\documentclass[superscriptaddress,aps,prapplied,footnoteinbib,reprint]{revtex4-2}
\usepackage{amsmath,amssymb,amsthm,hyperref,graphicx,bbold}
\hypersetup{hidelinks}
\usepackage[capitalise]{cleveref}
\hypersetup{colorlinks=true, linkcolor=blue, citecolor=red, urlcolor=blue}
\usepackage[usenames]{color}

\begin{document} 

\title{Non-linear dynamics near exceptional points of synthetic antiferromagnetic spin-torque oscillators}

\author{R. A. Duine}%

\affiliation{%
	Institute for Theoretical Physics, Utrecht University, 3584CC Utrecht, The Netherlands}%
\affiliation{Department of Applied Physics, Eindhoven University of Technology, P.O. Box 513, 5600 MB Eindhoven, The Netherlands}

\author{V. Errani}
\email{v.errani@uu.nl}
\thanks{Both authors contributed equally}

\affiliation{%
	Institute for Theoretical Physics, Utrecht University, 3584CC Utrecht, The Netherlands}%

\author{J. S. Harms}%
\email{j.harms@uu.nl}
\thanks{Both authors contributed equally}

\affiliation{Institute for Theoretical Physics, Utrecht University, 3584CC Utrecht, The Netherlands}%

\begin{abstract}
	We consider a synthetic antiferromagnetic spin-torque oscillator with anisotropic interlayer exchange coupling. This system exhibits exceptional points in its linearized dynamics. We find the non-linear dynamics and the dynamical phase diagram of the system both analytically and numerically. Moreover, we show that, near one of the exceptional points, the power of the oscillator depends extremely sensitively on the injected spin current. Our findings may be useful for designing sensitive magnetometers and for other applications of spin-torque oscillators. 
\end{abstract}
  
\maketitle
  
\section{Introduction}
For the design of many applications, such as magnetometers, converters and amplifiers, a strong response to perturbations is preferred.          
One way to achieve this strong response, is to make use of the existence of exceptional points (EPs) \cite{chen2017exceptional,wiersig2014enhancing,miri10science}.
EPs are characterized by a square-root dependence of the imaginary part of the eigenfrequecies on some system parameters, which enables a large dynamic response as a result of a small change in a parameter.
Mathematically, EPs correspond to the coalescence of different eigenvalues and eigenvectors in parameter space~\cite{heiss2012physics,heiss2012physics,dembowski2001experimental}.
EPs are studied intensely since they might lead to better sensors~\cite{chen2017exceptional,wiersig2014enhancing,miri10science}, and yield a variety of interesting phenomena such as lasing~\cite{feng2014single}, spontaneous emission~\cite{lin2016enhanced}, and give rise to geometric phases when encircling them~\cite{dembowski2004encircling}.
Examples of physical systems that exhibit EPs are optical microcavities \cite{chowdhury2022exceptional} and other photonic systems \cite{Miri2019,ozdemir2019parity}, optical lattices with engineered defects \cite{longhi2014optical}, electromechanical systems \cite{renault2019virtual}, superconducting resonators \cite{partanen2019exceptional}, nodal superconductors \cite{chowdhury2022exceptional}, semimetals \cite{gonzalez2017topological,molina2018surface,lee2015macroscopic} and  magnetic systems \cite{yang2018antiferromagnetism,yu2020higher,wang2021enhanced,jeffrey2021effect,liu2019observation,tserkovnyak2020exceptional,deng2022exceptional}.
While EPs have been studied intensely in the linear regime, the non-linear regime remains relatively unexplored.

In this article, we consider a synthetic antiferromagetic (SAF) spin-torque oscillator (STO), i.e., a spin-torque oscillator that consists of two magnetic layers that are coupled  by the Ruderman–Kittel–Kasuya–Yosida (RKKY) interactions. We consider the situation that one of the two magnetic layers is driven by means of the injection of spin current. This could be achieved by spin-orbit torque or by spin-transfer torque \cite{brataas2012}. 
Generically, an STO is a magnetic system in which the damping is compensated by the injection of spin angular momentum from a spin current to yield precessional magnetic dynamics \cite{kim2012spin, firastrau2013spin,houssameddine2007spin,deng2022exceptional}.
These oscillators have potential for a wide range of applications, such as detectors, microwave
signal sources \cite{tiberkevich2007microwave,chen2020spin}, microwave-assisted magnetic recording and neuromorphic computation \cite{chen2016spin,markovic2019reservoir}.
The SAF STO that we consider has anisotropic RKKY coupling and exhibits EPs in its linearized dynamics. We consider the full non-linear dynamics analytically and find the limit cycles of the magnetization dynamics. Moreover, we show that the dynamics becomes relatively simple because the power and precessing frequency depend linearly on the injected spin current close to the EPs. These analytical results agree with numerical computations and lead to a complete understanding of the steady-state behaviour of the system.
Furthermore, we find from our analysis that the magnetization dynamics is extremely sensitive to small changes in parameters in the sense that the slope of the steady-state power with respect to injected spin current diverges close to one of the EPs. This result can not be obtained from a linear analysis and the non-linear description is therefore needed. A complementary work to ours is that of Deng {\it et al.} \cite{https://doi.org/10.48550/arxiv.2205.02308} who numerically consider the non-linear dynamics of an STO near a different type of EP. Another recent closely-related complementary work is that of Ref.~\cite{https://doi.org/10.48550/arxiv.2209.01572}.

This remainder of this article is organized as follows, in the first section, we present the model and determine the eigenfrequencies and stability conditions from a linearized analysis. In the second section, we consider the non-linear dynamics close to the EPs and find the limit circles of the magnetic dynamics. 
In the third section, we consider the magnetization dynamics numerically and find the dynamical phase diagram. We end with a brief conclusion and outlook.

\section{Linear Analysis}

\label{sec:physical-model}
\begin{figure}[h]
	\centering
	\includegraphics[width=.9\columnwidth]{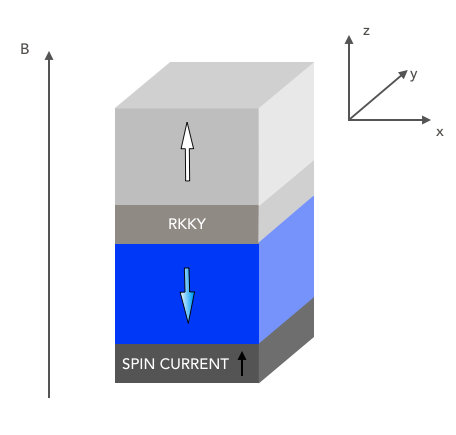}
	\caption{ A synthetic antiferromagnetic spin torque
oscillator in an external magnetic field $B \hat{z}$. Spin angular momentum is injected into the bottom magnetic layer.  
	}
	\label{fig:2Spinspic}
\end{figure}

We consider a spin-torque oscillator composed of two RKKY coupled nanomagnets subject to the same external magnetic field and uniaxial anisotropy, see~\cref{fig:2Spinspic}. For simplicity, we take both magnetic layers to be identical.
Furthermore, spin angular momentum is injected into the bottom magnetic layer. The magnetic energy is given by
\begin{align}\label{eq:Hamiltonian}
	E
	=&-B(m_{U,z}+m_{L,z})-K ({m}_{U,z}^2+ {m}_{L,z}^2)/2
	\\\nonumber
	&-J_\perp(m_{U,x} m_{L,x}+ m_{U,y} m_{L,y}) - J_z m_{U,z} m_{L,z}~,
\end{align}
where $ U $ denotes the upper- and $ L $ the lower macrospin, and $m_{L/U,i}$ the $i$-th Cartesian component of the macrospin. 
Furthermore, $K$ is the uniaxial anisotropy constant, $ B>0 $ the external magnetic field directed along the $ z $ axis, $ J_\perp $ the in-plane RKKY interaction, and $J_z$ its out-of-plane component. We take the RKKY interaction to be anisotropic because, first of all, it will typically be anisotropic, and second, this enriches the phase diagram.

For temperatures below the Curie temperature the magnetization dynamics is well described by the Landau-Lifschitz-Gilbert (LLG) equation with the inclusion of the injected spin current
\begin{equation}\label{LLG}
	\frac{\partial \mathbf{m}_\nu}{\partial t}=-\mathbf{m}_\nu \times \mathbf{h}_{\mathrm{eff},\nu} + \alpha \mathbf{m}_\nu \times \frac{\partial \mathbf{m}_\nu}{\partial t} + I_{s,\nu} \mathbf{m}_\nu \times (\mathbf{m}_\nu\times\hat{z}),
\end{equation}
with the effective field $ \mathbf{h}_\mathrm{eff,\nu}=-\gamma\delta E/\delta\mathbf{m}_\nu $.
Here, $ \nu $ denotes either the lower $ (L) $ or the upper $ (U) $ macrospin, $ \alpha $ the dimensionless Gilbert damping and $ I_{s,U}=0 $, $ I_{s,L}= I_s>0 $ the spin current. The sign of the spin current is such that it tends to align the bottom magnetic layer against the external field. Since the Gilbert damping is typically small, $\alpha \ll 1$, we work in most of what follows to lowest order in $\alpha$ and discard terms of the order $\alpha I_s$ as well. Furthermore, we set $\gamma=1$ so that $B,K, J_\perp$ and $J_z$ have units of frequency. 

For the case that $K>0$, the magnetic energy (\ref{eq:Hamiltonian}) yields four configurations where the torques on the magnetization direction in both layers vanish simultaneously: these are both layers pointing up, both pointing down, and the two anti-parallel configurations. Depending on the parameters, these configurations are energy minima that are stable, or unstable. For large fields, the configuration with both spins pointing up is stable. For large spin currents, the bottom layer is forced to point downward, while, depending on the strength and sign of the RKKY coupling, the top layer may point down or up. 

The most interesting configuration for our purposes is the antiparallel configuration. As we shall see below, the linearized dynamics around this configuration yields an EP. This may be anticipated because, in the absence of dissipation, i.e., when $\alpha=0$ and $I_s=0$, the antiparallel configuration in an external field is reminiscent of a system of two coupled harmonic oscillators with one of the oscillators having a potential energy that is inverted. This latter system is known to yield an EP \cite{PhysRevA.98.023841}. Because $I_s>0$ we consider the situation where the bottom magnetic layer is pointing against the field.

We investigate the stability of a given magnetic state by linearizing the LLG equations for small deviations around that state. For the reasons mentioned above, we focus on the antiparallel configuration with the magnetization of the bottom layer pointing against the external field, which yields the eigenfrequencies
\begin{align}\label{eq:eigenfreq-antiparallel}
	&(\alpha^2+1)\omega_\pm
	=
	{B-\mathrm{i}\alpha(K-J_z)}
	-(\mathrm{i}-\alpha)I_s/2 \nonumber 
	\\
	&\pm
	\sqrt{[K-J_z- \mathrm{i} \alpha  B+(\mathrm{i}-\alpha)I_s/2]^2-(\alpha ^2+1) J_\perp^2}~.
\end{align}
From this expression, we find that the parameters for which the system exhibits an EP by setting the expression under the square root equal to zero. To lowest order in $\alpha$ and $I_s$ this yields $J_\perp^2=(K-J_z)^2$ and $I_s=2\alpha B$. Close to this EP and, in particular,  when the expression underneath the square-root is negative, the imaginary part of the eigenvalues depends strongly on small changes in parameters. Physically, this implies that a small change in parameters may yield a strong dynamic response, because a positive imaginary part of the eigenfrequency corresponds to exponential growth of small-amplitude fluctuations.  By determining when the imaginary part of the above eigenfrequency changes sign,  we find that the antiparellel configuration is stable when
\begin{subequations}\label{eq:stability-condition-antiparallel} 
	\begin{align}
		\frac{I_s}{2\alpha}
		&>
		\frac{
			-K-J_z+{|K-J_z|B}/{\sqrt{(K-J_z)^2-J_\perp^2}
			}
		}{
			1+{|K-J_z|}/{\sqrt{(K-J_z)^2-J_\perp^2}}
		},
		\\
		\frac{I_s}{2\alpha}
		&<
		\frac{
			-K-J_z-{|K-J_z|B}/{\sqrt{(K-J_z)^2-J_\perp^2}
			}
		}{
			1-{|K-J_z|}/{\sqrt{(K-J_z)^2-J_\perp^2}}~.
		}.
	\end{align}
\end{subequations}
In the next section we focus on the non-linear dynamics outside this range of dynamical stability. 
As we shall see, the power of the STO is extremely sensitive to small changes in the current around this EP for $ K>0 $, but not when $ K<0 $. 

\section{Non-linear dynamics}\label{sec:non-linear-dynamics}
In this section we discuss the non-linear dynamics of the system described in Section~\ref{sec:physical-model} and in specifically focus on the limit cycles of the model. 

The reactive dynamics is formulated by means of the Poisson bracket
$ \{m_{\alpha},m_{\beta}\}=\epsilon_{\alpha\beta\gamma}m_\gamma $.
From here we define the canonical coordinates $ p_{U/L}=m_{U/L,z} $ and $ \theta_{U/L}=\arctan(m_{U/L,y}/m_{U/L,x}) $, that correspond to the total power and angle of the oscillator and which have non-zero Poisson brackets $ \{p_{U/L},\theta_{U/L}\}=1 $. 
We continue performing yet another coordinate transformation that makes use of the rotational symmetry around the $z$-axis
\begin{align}
	\begin{array}{ll}
		\mu=(p_U+p_L)/2,&\eta=(p_U-p_L)/2,\\
		\theta=\theta_{U}+\theta_{L},&\phi=\theta_{U}-\theta_{L},
	\end{array}
\end{align}
where $ \mu\in[-1,1] $ and $ \eta\in[-1+\mu,1-\mu] $.
The above coordinates have $ \{\mu,\theta\}=1 $ and $ \{\eta,\phi\}=1 $ as non-vanishing Poisson brackets.
In these coordinates the   Hamiltonian (internal energy) in~\cref{eq:Hamiltonian} becomes
\begin{align}
	h\equiv E
	=&-2B\mu-K(\mu^2+\eta^2)-J_z(\mu^2-\eta^2)\\\nonumber
	&-J_\perp\sqrt{[1-(\mu+\eta)^2][1-(\mu-\eta)^2]}\cos(\phi),
\end{align}
where the rotation symmetry around the $z$-axis ensures that the right-hand side of the above does not depend on $\theta$. The Hamilton equations of motion are accordingly given by
\begin{subequations}
	\begin{align}
		\dot{\mu}=&
		\{\mu,E\}=\{\mu,\theta\}\partial_\theta E=
		0,\\
		\dot{\theta}
		=&2B+2(K+J_z)\mu
		\\\nonumber&
		-\frac{2J_\perp \mu(1+\eta^2-\mu^2)\cos(\phi)}{\sqrt{[1-(\mu+\eta)^2][1-(\mu-\eta)^2]}},\\\label{eq:dot-phi}
		\dot{\phi}
		=&2(K-J_z)\eta-\frac{2J_\perp\eta(1+\mu^2-\eta^2)\cos(\phi)}{\sqrt{[1-(\mu+\eta)^2][1-(\mu-\eta)^2]}},\\\label{eq:dot-eta}
		\dot{\eta}
		=&2J_\perp\sqrt{[1-(\mu+\eta)^2][1-(\mu-\eta)^2]}\sin(\phi),
	\end{align}
\end{subequations}
where the rotation symmetry guarantees that the total power is conserved. For the limit cycles we expect $ \dot{\phi}=\dot{\eta}=0 $, since both spins experience the same magnetic field strength and are thus expected to have equal angular velocity.
Using this ansatz, we find two possible expression for the difference in power $ \eta $ in terms of the total power $ \mu $ for $ J_\perp^2<K^2 $, which are given by
\begin{align}\label{eq:eta-mu-relation}
	\eta&=0,~
	\eta^2=1+\mu^2-\frac{2|K-J_z||\mu|}{\sqrt{(K-J_z)^2-J_\perp^2}}.
\end{align}
In case $ J_\perp^2>(K-J_z)^2 $ we are only left with $ \eta=0 $ as a solution, since we require $ \eta $ to be real.

The model under consideration furthermore has dissipative contribution in the form of Gilbert damping and the injection of spin angular momentum. Up to first order in the Gilbert damping constant $ \alpha $ the dynamics of total power $ \mu $ and the relative power $\eta$ due to Gilbert damping is given by
\begin{subequations}
	\begin{align}\label{eq:dissipation-Gilbert}
		\dot{\mu}/\alpha=&
		\nonumber
		\left[B+(K+J_z)\mu\right](1-\mu^2-\eta^2)
		-2(K-J_z)\eta^2\mu\\
		-&J_\perp\sqrt{[1-(\mu+\eta)^2][1-(\mu-\eta)^2]}\cos(\phi)\mu
		\\\nonumber
		=&
		B(1+\mu^2-\eta^2)+[h+J_z+K(1-2\eta^2)]\mu.
		\\
		\dot{\eta}/\alpha=&
		\nonumber
		\big[-2B\mu+K(1-3\mu^2-\eta^2)-J_z(1+\mu^2-\eta^2)\\
		-&J_\perp\sqrt{[1-(\mu+\eta)^2][1-(\mu-\eta)^2]}\cos(\phi)\big]\eta.
		\\\nonumber
		=&
		[h+K(1-2\mu^2)-J_z]\eta.
	\end{align}
\end{subequations}
The influence of SOT, up to first order in $ I_s $, on the dissipative dynamics is given by
\begin{subequations}
	\begin{align}\label{eq:dissipation_SOT}
		\dot{\mu}/I_s=&
		[(\mu-\eta)^2-1]/2,
		\\
		\dot\eta/I_s=&
		[1-(\mu-\eta)^2]/2.
	\end{align}
\end{subequations}
With this set up we are in the position to discuss the limit cycles of the model and their stability.
Again for all values of $ J_\perp $, $ \eta=0 $ will be a solution of $ \dot{\phi}=\dot{\eta}=0 $. Furthermore, for $ J_\perp^2<K^2 $ we additionally have  $ \eta^2=1+\mu^2-{2|K||\mu|}/{\sqrt{K^2-J_\perp^2}} $ as a solution. Below we address limit cycles and stability in for the solution $ \eta=0 $, which covers, as we shall see, the behaviour near the EP. For completeness, we address the other type of limit cycles and their stability in~\cref{app:further-limit-cycle,app:stability-eta-0,app:stability-eta-neq-0,app:phase-diagram}.

When $ \eta=0 $ the equation for the total power $ \mu $ becomes
\begin{align}\label{eq:fixed-point-mu-eta=0}
	\dot{\mu}/\alpha=&[B-I_s/2\alpha
	+(K+J_z
	-|J_\perp|)\mu](1-\mu^2).
\end{align}
This has two solutions
\begin{align}\label{eq:limit-cycle-mu-sol}
	\mu=\pm1,~\mu=\frac{B-I_s/2\alpha}{|J_\perp|-J_z-K},
\end{align} 
with $ \mu=m_{U,z}=m_{L,z} $.
The first solution describes both macrospins aligning in the $ \pm \hat{z} $ direction.
The second solution on the other hand describes the limit cycle towards which the system converges for sufficiently large times.
Furthermore, this limit cycle has a precessional frequency of
\begin{equation}\label{eq:precessional-frequency}
	\dot{\theta}=I_s/\alpha.
\end{equation}
From the fixed point analyses presented in~\cref{app:stability-eta-0} these fixed points are stable if $ \partial_\mu\dot{\mu}<0 $ and $ |J_\perp|(1+\mu^2)>(K-J_z)(1-\mu^2) $.
Alternatively, these limit cycles are unstable if one of the above constraints is not satisfied.
The limit cycle solution $ \mu={(B-I_s/2\alpha)}/{(|J_\perp|-J_z-K)} $ is thus stable if
$
	\max\left[K+J_z,(K-J_z){(1-\mu^2)}/{(1+\mu^2)}\right]<|J_\perp|.
$
While the static solution with $ \mu=\pm1 $ on the other hand is stable if
$
	\mp(B-I_s/2\alpha)<K+J_z-|J_\perp|.
$

In conclusion, we find that the power of the SAF is quite sensitive to small perturbations around the point $ |J_\perp|\gtrsim K+J_z $, where the slope of the total power $ \mu=m_{U,z}=m_{L,z} $ with respect to $ I_s/2\alpha $ is given by $ (|J_\perp|-J_z-K)^{-1} $ in~\cref{eq:limit-cycle-mu-sol}. 
The total power of the oscillator therefore depends sensitively on injected spin current around this point. 
Hence, we find an enhanced sensitivity if $ K+J_z>0 $ for $ J_\perp>K+J_z $.

\section{Numerical Results}\label{sec:phase-diagram}
\begin{figure}[t]
	\includegraphics[width=\columnwidth]{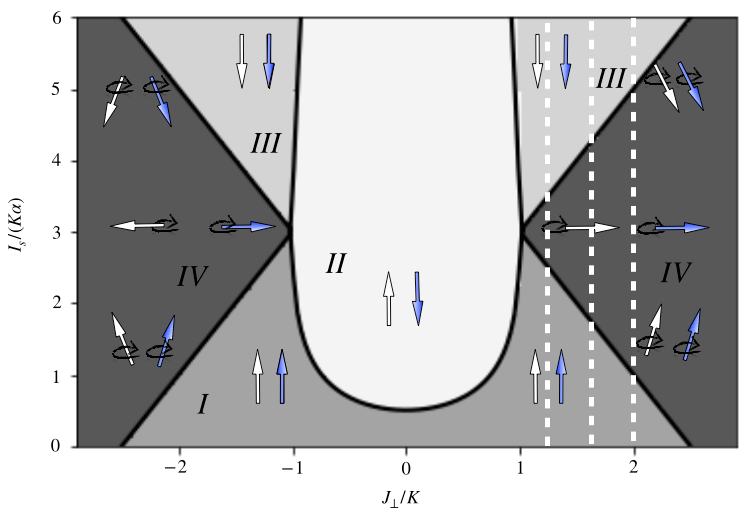}
	\caption{\label{fig:phase-diagram}
	Dynamical phase diagram for $K>0$ as a  function of coupling $J_\perp$ and current $I_s$. The analytical predictions from \cref{eq:stability-condition-antiparallel,eq:limit-cycle-mu-sol} are plotted as the black lines. The steady-state configuration is indicated for each region. The region IV correspond to oscillations. The $z$-component of the magnetization in this region depends on the injected spin current. We took $B/K=1.5$ and $J_z=0$.
	}
\end{figure}
\begin{figure}[t]
	\centering
	\includegraphics[width=\columnwidth]{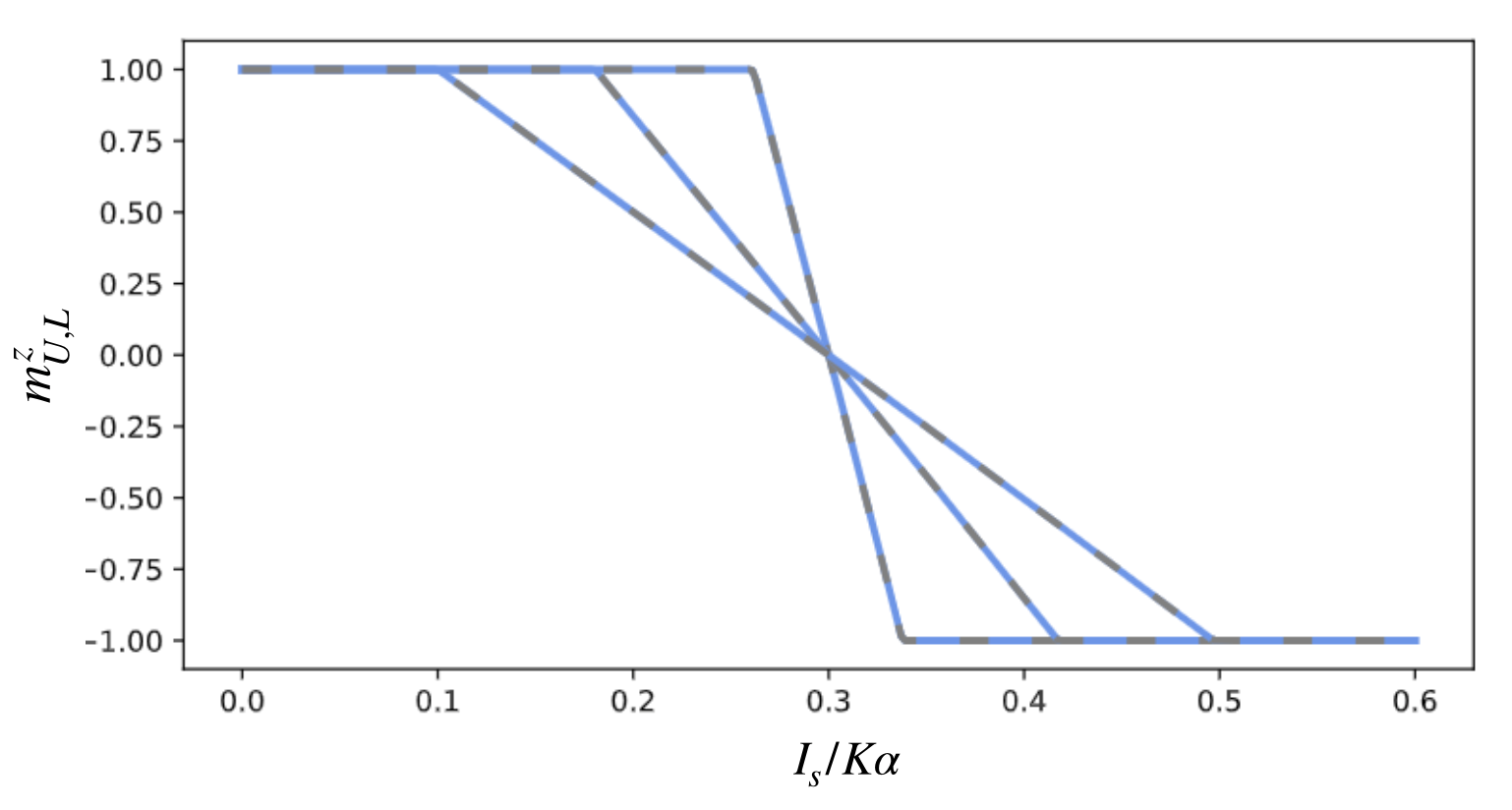}
	\caption{ Plot of the $ z  $ direction of the macrospins in both synthetic layers as a function of the current $I_s$. The three different lines correspond to different values in the coupling strength $J_\perp$, with the steeper slope corresponding to decreasing $J_\perp$. These values are depicted by the dotted lines in~\cref{fig:phase-diagram}. Parameters taken are $B/K=1.5$ and $J_z=0$.}
	\label{fig:mu-vs-current}
\end{figure}
In this section we determine all the dynamical phases of the synthetic antiferromagnetic oscillator using numerical solutions.
For concreteness, we focus on the case that $J_z=0$. We solve the LLG equation~\eqref{LLG} numerically for different values of the coupling $J_\perp$ and current $I_s$ starting from the antiparallel configuration with a small initial perturbation.
In \cref{fig:phase-diagram} we present the long-time behaviour in each region of the dynamical phase diagram, with the blue spin denoting the magnetic layer that is driven by spin current. We note that a change in the external magnetic field $B$ results in a rigid upward- or downward shift of the phase diagram. 
The black lines in~\cref{fig:phase-diagram} are the analytically predicted phase boundaries from~\cref{eq:stability-condition-antiparallel,eq:limit-cycle-mu-sol}.
We find a very good agreement between the analytic predictions and the results from numerical solutions.

There are three regions where the spin configuration is static (I, II, and III), and one region, IV, where the magnetization is oscillating. The region in~\cref{fig:phase-diagram} with $ J_\perp<K $ may be interpreted in the following way.
If we start from zero coupling and increase the current we observe that the configuration is initially parallel with both macrospin aligned upward (region I).
This is what we expect since for small values of the current the spin current is not large enough to compensate the Gilbert damping and the two macro-spins both align with the external magnetic field.
Increasing the current further, the system is able to keep the initial configuration (region  II), since now the current is enough to compensate for the damping (light-blue macro-spin pointing downwards), but the coupling is too small to make the white macrospin, i.e., the magnetic layer into which no spin current is injected, flip.
Indeed for increasing values of the coupling and sufficiently high current both macro-spins are aligned against the external field (region  III).

When considering the non-linear behaviour beyond the EP, region IV is of most interest. In this region the two macro-spins are oscillating, with a frequency $I_s/\alpha$ depending only on the current $I_s$ and the damping $\alpha$. These two macrospins exhibit nearly parallel orientations for positive values of the coupling and antiparallel orientations for negative values. 
In~\cref{fig:mu-vs-current} we show the $ z $ component of the magnetization as a function of the current $ I_s $, in which the different lines correspond to the vertical dashed lines in \cref{fig:phase-diagram}. The numerical simulations confirm that the current and the macrospin orientation are related in a linear way given by~\cref{eq:limit-cycle-mu-sol}. 

We have focused on the situation that $J_z=0$, which is the most interesting because all the four phases (I, II, III, IV) meet at one point. For the case that $0<J_z \neq J_\perp$, we find that this point splits into two points, one where I, II, III meet, and one where I, III, IV meet. The distance between these two points scales with $J_z$. Near the point where regions I, III, IV meet, we again have that the slope of the power of the oscillator with spin current diverges upon approaching this point.

\section{Discussion and Conclusions}
We analytically and numerically explored the non-linear behaviour of a synthetic antiferrmagnetic spin-torque oscillator with anisotropic RKKY coupling. We found that the EP which is found in its linearized dynamics leads to enhanced sensitivity of the power of the oscillator, in particular as a function of injected spin current. This enhanced sensitivity may be used to engineer magnetometers or sensors of spin current. Furthermore, recent works shows that it is possible to use 
spin torque nano-oscillators to implement different computing schemes and to classify waveforms~\cite{markovic2019reservoir}. Moreover, spin-torque oscillators may be used as tunable spin-wave emitters that exite a specific spin wave depending on the current. We expect that the enhanced sensitivity we predict is an asset for such applications. 

Regarding the experimental realization of our model, an ingredient is the anisotropic interlayer coupling.
This has been proposed and experimentally observed in Refs.~\cite{xia1997noncollinear,li2008oscillatory}. The required anisotropies may be engineered by interfaces with heavy metals and/or engineering the shape of the magnetic layers. 

Future work could focus on the inclusion of thermal fluctuations. We expect  that these will affect the phase but not the orientation or the amplitude of the oscillations. Finally, we hope that our work, together with that of Deng {\it et al.} \cite{https://doi.org/10.48550/arxiv.2205.02308}, inspires the operation of STOs near EPs

\section*{Acknowledgements}
R. A. Duine conceived the project, V. Errani and J. S. Harms performed the linear analysis, V. Errani wrote the numerical codes and numerically constructed the phase diagram and J. S. Harms performed the analytic non-linear analysis.  V. Errani and J.S. Harms wrote the first draft after which all co-authors contributed to finalizing it. 
J. S. Harms would like to thank W. Q. Boon for fruitful discussions.
R. A. Duine is member of the D-ITP consortium, a program of the Netherlands Organisation for Scientific Research (NWO) that is funded by the Dutch Ministry of Education, Culture and Science (OCW). R.A.D. acknowledges the funding from the European Research Council (ERC) under the European Union's Horizon 2020 research and innovation programme (Grant No. 725509).
This work is part of the Fluid Spintronics research programme with project
number 182.069, which is financed by the Dutch
Research Council (NWO).
\appendix
\section{Second type of limit cycles for $ J_\perp^2<K^2 $ for $ J_z\rightarrow0 $}\label{app:further-limit-cycle}
For $ J_\perp^2<K^2 $ we also need to consider $ \eta^2=1+\mu^2-{2|K||\mu|}/{\sqrt{K^2-J_\perp^2}} $ as a solution to $ \dot{\phi}=\dot{\eta}=0 $.
Using the above ansatz we find that~\cref{eq:dissipation-Gilbert,eq:dissipation_SOT} give
\begin{align}\label{eq:eqom-limit-cycle-J_perp<K}
	&\dot{\mu}/\alpha=\\\nonumber
	&\left[B+ K\mu-\frac{I_s}{2\alpha}\right]
	\left[-2\mu^2+\frac{2|K||\mu|}{\sqrt{K^2-J_\perp^2}}\right]
	\\\nonumber
	&-\mu\frac{J_\perp^2}{K}\frac{2|K||\mu|}{\sqrt{K^2-J_\perp^2}}
	\\\nonumber
	&-2K\mu\left[1+\mu^2-\frac{2|K||\mu|}{\sqrt{K^2-J_\perp^2}}\right]
	\\\nonumber
	&-\frac{I_s}{\alpha}\mathrm{sgn}(\eta)\mu\sqrt{1+\mu^2-\frac{2|K||\mu|}{\sqrt{K^2-J_\perp^2}}}.
\end{align}
In first instance we note that $ \mu=0 $ gives a static solution to the above equation.
This describes the configuration in~\cref{fig:2Spinspic}, in which two spins point in opposite direction.
From~\cref{app:stability-eta-neq-0} it follows that a requirement for stability of this phase is $ \partial_\mu\dot{\mu}<0 $ at the fixed point.
This requirement for stability, with $ \mathrm{sgn}(\eta)=1 $, becomes
\begin{subequations}\label{eq:stability-requirement-static-appendix}
\begin{align}
	\frac{I_s}{2\alpha}&>\frac{-K+|K|B/\sqrt{K^2-J_\perp^2}}{\;\,1+|K|/\sqrt{K^2-J_\perp^2}},\\
	\frac{I_s}{2\alpha}&<\frac{-K-|K|B/\sqrt{K^2-J_\perp^2}}{\;\,1-|K|/\sqrt{K^2-J_\perp^2}}.
\end{align}
\end{subequations}
These are precisely the same conditions as in~\cref{eq:stability-condition-antiparallel} for $ J_z\rightarrow0 $.
The fixed point with $ \mathrm{sgn}(\eta)=-1 $ on the other hand is stable if
\begin{subequations}
\begin{align}
	\frac{I_s}{2\alpha}&>\frac{\;-K+|K|B/\sqrt{K^2-J_\perp^2}}{-1+|K|/\sqrt{K^2-J_\perp^2}},\\
	\frac{I_s}{2\alpha}&<\;\;\;\frac{\;K+|K|B/\sqrt{K^2-J_\perp^2}}{1+|K|/\sqrt{K^2-J_\perp^2}}.
\end{align}
\end{subequations}
we see that the configuration with $ \mathrm{sgn}(\eta)=-1 $ is always unstable, since there is no interval for stability. 

In order to make progress in the regime where we expect limit cycles we rewrite~\cref{eq:eqom-limit-cycle-J_perp<K}

\begin{align}\label{eq:eq-fixed-point-eta-nonzero}
	g(I_s,\mu)&=
	\dot{\mu}/\alpha|\mu|
	\\&=\nonumber
	\left(1-2\epsilon|\mu|\right)
	\\&\nonumber
	\times\bigg[
	2K[\mu-\mathrm{sgn}(\mu)\epsilon]
	+\bigg(
	B-\frac{I_s}{2\alpha}
	\bigg)
	\bigg]
	\\&\nonumber
	-\mathrm{sgn}(\mu)\epsilon\frac{I_s}{\alpha}\eta(\mu),
\end{align}
where $ \epsilon\equiv\sqrt{K^2-J_\perp^2}/2|K|\in(0,1/2) $. 
Let us analyse te above equation in a bit more detail.
We assume that $ \mu>0 $ at the fixed point, from here we find that~\cref{eq:eq-fixed-point-eta-nonzero} implies 
\begin{equation}\label{eq:current-of-mu}
	\frac{I_s}{2\alpha}
	=
	\frac{B+2K(\mu-\epsilon)}{1+2\epsilon\eta/(1-2\epsilon\mu)}.
\end{equation}
From~\cref{app:stability-eta-neq-0} we find the stability requirement $ \dot{\mu}\leq0 $ for limit cycles of~\cref{eq:eq-fixed-point-eta-nonzero} to be
\begin{align}\label{eq:stability-limit-cyle-eta-nonzero}
	\frac{I_s}{2\alpha}\frac{1-4\epsilon^2}{(1-2\epsilon\mu)^2}
	+2K\eta
	<&0.
\end{align}
We like to find the minimal current $ I_s $ for these limit cycles to be stable, hence we like to find the current for which~\cref{eq:stability-limit-cyle-eta-nonzero} is zero, in other words $ \partial_\mu\dot\mu=\partial_\mu g(I_s,\mu)\rvert_{I_s(\mu)}=0 $.
On the other hand we know $ g(I_s(\mu),\mu)=0 $ is the limit cycle --fixed point-- condition.
From here we find $ d_\mu g(I_s(\mu),\mu)=(\partial_{I_s}g) (\partial_\mu I_s)+\partial_\mu g=0 $, and hence $ \partial_\mu I_s=0\leftrightarrow\partial_\mu g\equiv\partial_\mu\dot\mu=0. $
Thus when looking for the critical current it is sufficient to consider $ \partial_\mu I_s=0 $.
We proceed by writing~\cref{eq:current-of-mu} as
\[
(1-2\epsilon\mu+2\epsilon\eta)I_s/2\alpha=(1-2\epsilon\mu)(B+2K(\mu-\epsilon)). 
\]
Furthermore, we take the derivative with respect to $ \mu $ of the above and note that we are considering points which satisfy $ \partial_\mu I_s=0 $.
This leaves us with
\[
(1-2\epsilon\mu+2\epsilon\eta)I_s/2\alpha
=
[2\epsilon(B+2K(\mu-\epsilon))+2K(1-2\epsilon\mu)]\eta.
\]
which according to~\cref{eq:current-of-mu} gives us
\begin{equation}\label{eq:equation-critical-value-mu}
\begin{aligned}
	&(1-2\epsilon\mu)(B+2K(\mu-\epsilon))
	\\=&
	[2\epsilon B-2K(1-4\epsilon\mu+2\epsilon^2)]\eta.
\end{aligned}
\end{equation}
In order to proceed we assume $ \eta^2\ll1 $ around the critical current.
A consequence of the above is that we assume $ \mu_\epsilon-\delta\mu\ll1 $ with
\begin{equation}
\mu_\epsilon=\frac{1}{2\epsilon}\left(1-\sqrt{1-4\epsilon^2}\right),
\end{equation}
and from relation~(\ref{eq:eta-mu-relation}) we see that $ \eta^2\sim(2\epsilon)^{-1}\sqrt{1-4\epsilon^2}\,\delta\mu $ at linear order in $ \delta\mu $  and $ \eta^2 $. By squaring~\cref{eq:equation-critical-value-mu} and expanding up to linear order in $ \delta\mu $ we find the solution to be
\begin{equation}
	2\delta\mu
	\sim
	\sqrt{\frac{1}{4\epsilon^2}-1}
	\frac
	{\left[K\big(1-2\epsilon^2-\sqrt{1-4\epsilon^2}\big)\;\,+\epsilon B\right]^2}
	{\left[K\big(1-2\epsilon^2-2\sqrt{1-4\epsilon^2}\big)+\epsilon B\right]^2}.
\end{equation}
Accordingly the critical current is well approximated by using~\cref{eq:current-of-mu}
\begin{equation}\label{eq:critical-current-histeretic-regime}
	I_{s,c}\simeq I_s(\mu_\epsilon-\delta\mu).
\end{equation}
We thus have two forms of stable limit cycles for $ -|J_\perp|<K<0 $ and $ I_s>I_{s,c} $, namely one described in~Section~\ref{sec:non-linear-dynamics} where $ \eta=0 $ and one with $ \eta\neq0 $ which we described in this Appendix.
\section{Stability requirements for limit cycles with $ \eta=0 $}\label{app:stability-eta-0}
To perform the stability analyses, we first determine the fixed points in $ \phi $ and $ \eta $ up to first order in dissipative terms. Including dissipative corrections~\cref{eq:dot-eta} become
	\begin{align}
		\delta\dot{\eta}
		&=
		4|J_\perp|(1-\mu^2)\delta\phi_\mathrm{fp}-\frac{I_s}{2}(1-\mu^2)=0.
	\end{align}
The above implies 
$ \delta\phi_\mathrm{fp}=I_s/8|J_\perp| $.
We continue the stability analyses by linearizing around the fixed point $ \eta=0 $, $ \sin(\phi)=I_s/4J_\perp $ and $ \mu=(B-I_s/2\alpha)/(|J_\perp|-K) $,
\begin{align}
	\begin{pmatrix}
		\dot{\phi}\\
		\dot{\eta}\\
		\dot{\mu}
	\end{pmatrix}
	=
	\begin{pmatrix}
		0&\gamma^\phi_\eta&0\\
		\gamma^\eta_\phi&\epsilon^\eta_\eta&0\\
		0&\epsilon^\mu_\eta&\epsilon^\mu_\mu		
	\end{pmatrix}
	\begin{pmatrix}
		\delta\phi\\
		\delta\eta\\
		\delta\mu
	\end{pmatrix}.
\end{align}
Where $ \gamma $ is zeroth order in $ \alpha $ and $ I_s $, and $ \epsilon $ is first order in dissipation.
The eigenvalues $ \lambda $ of the above matrix are given by
\begin{equation}
	(\epsilon^\mu_\mu-\lambda)
	[(\epsilon^\eta_\eta-\lambda)\lambda-\gamma^\eta_\phi\gamma^\phi_\eta]
	=0.
\end{equation}
Hence, one eigenvalue is given by $ \epsilon^\mu_\mu $ and the real part of the other two eigenvalues is given by $ \epsilon^\eta_\eta $. The fixed point is stable if the real part of all eigenvalues is negative. 
The constraint $ \epsilon^\mu_\mu<0 $ precisely gives $ \partial_\mu\dot{\mu}<0 $ in~\cref{eq:fixed-point-mu-eta=0}. The requirement that $ \epsilon^\eta_\eta<0 $ on the other hand gives
$
		\epsilon^\eta_\eta/\alpha
	=
	(K-J_z)(1-\mu^2)-|J_\perp|(1+\mu^2)
	<0,
$
which is satisfied for $ K-J_z<|J_\perp| $.
\section{Stability requirements for limit cycles with $ \eta\neq0 $}\label{app:stability-eta-neq-0}
In case $ \eta\neq0 $, the angle $ \phi $ shifts due to dissipative corrections on~\cref{eq:dot-eta}.
We denote this shift by $ \delta\phi_\mathrm{fp} $.
We find that the linearized equations of motion around the fixed point give
\begin{align}
	\begin{pmatrix}
		\dot{\phi}\\
		\dot{\eta}\\
		\dot{\mu}
	\end{pmatrix}
	=
	\begin{pmatrix}
		\epsilon^\phi_\phi&\gamma^\phi_\eta&\gamma^\phi_\mu\\
		\gamma^\eta_\phi&\epsilon^\eta_\eta&\epsilon^\eta_\mu\\
		0&\epsilon^\mu_\eta&\epsilon^\mu_\mu		
	\end{pmatrix}
	\begin{pmatrix}
		\delta\phi\\
		\delta\eta\\
		\delta\mu
	\end{pmatrix}.
\end{align}
With $ \gamma $ zeroth order in $ \alpha $ and $ I_s $, and $ \epsilon $ first order in dissipation.
The eigenvalues $ \lambda $ of the above matrix are given by the third order polynomial equation
\begin{equation}
\begin{aligned}
	&(\epsilon^\mu_\mu-\lambda)
	[(\epsilon^\phi_\phi-\lambda)(\epsilon^\eta_\eta-\lambda)-\gamma^\eta_\mu\gamma^\phi_\eta]
	\\&
	+\gamma^\eta_\mu\gamma^\phi_\mu\epsilon^\mu_\eta
	-\epsilon^\mu_\mu\epsilon^\eta_\mu(\epsilon^\phi_\phi-\lambda)
	=0.
\end{aligned}
\end{equation}
Hence, the eigenvalues -- up to first order in $ \alpha $ and $ I_s $ are given by
$
\lambda_1
=
\epsilon^\mu_\mu-(\gamma^\phi_\mu/\gamma^\phi_\eta)\epsilon^\mu_\eta 
$
and
$ 
\lambda_\pm
=
(\epsilon^\phi_\phi+\epsilon^\eta_\eta)/2
\pm\sqrt{\gamma^\eta_\phi\gamma^\phi_\eta}
\pm\epsilon^\mu_\eta{\gamma^\eta_\phi\gamma^\phi_\eta}/
		{\sqrt{\gamma^\eta_\phi\gamma^\phi_\eta}}
$.
Since we're only interested in the real part of the eigenfrequencies for our stability analyses, we find that the limit cycle is stable if 
\begin{subequations}
\begin{align}
	\mathrm{Re}(\lambda_1\;)
	&=
	\epsilon^\mu_\mu-(\gamma^\phi_\mu/\gamma^\phi_\eta)\epsilon^\mu_\eta
	\\&\nonumber
	=\epsilon^\mu_\mu+(\partial_\mu\eta)\epsilon^\mu_\eta
	=\partial_\mu\dot{\mu}\leq0,
	\\
	\mathrm{Re}(\lambda_\pm)
	&=
	(\epsilon^\phi_\phi+\epsilon^\eta_\eta)/2\leq0.
\end{align}
\end{subequations}
First we note that $ (\epsilon^\phi_\phi+\epsilon^\eta_\eta)/2=(\mu-\eta)I_s/2\alpha-\mu(B+2K(\mu-\epsilon))\leq0 $, which implies $ \mathrm{sgn}[B+2K(\mu-\epsilon)]\eta>0 $.
This condition is satisfied at the limit cycle.
Hence we're left with the stability condition $ \partial_\mu\dot{\mu}\leq0 $.
\section{Phase diagram for $ K<0 $ with $ J_z\rightarrow0 $}\label{app:phase-diagram}
\begin{figure}
	\centering
	\includegraphics[width=\columnwidth]{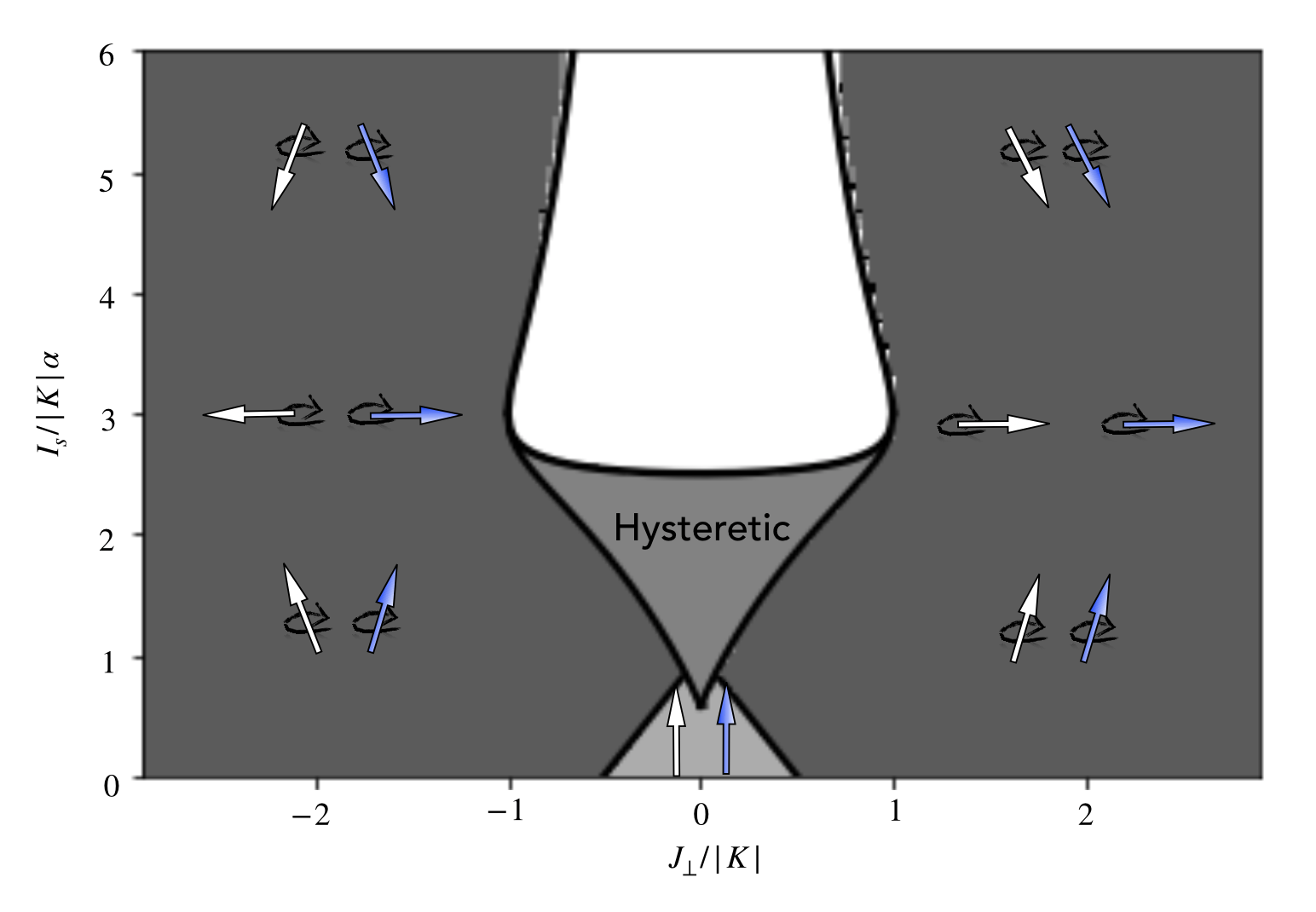}
	\caption{ Numerical phase diagram for the case $K<0$ as a function of coupling $J_\perp$ and current $I_s$. In black the analytical predictions from~\cref{eq:stability-condition-antiparallel,eq:stability-requirement-static-appendix,eq:critical-current-histeretic-regime}  are plotted. In each region the corresponding long time configuration is indicated. The  regions for large and small $J_\perp$ correspond to oscillations. The $z$-component of the magnetization in this region depends on the injected spin current. The characteristic of the hysteretic region is the dependence of the long-time configuration on the initial conditions. We took $B/K=1.5$ and $J_z=0$. }
	\label{fig:phase-diagram-K<0}
\end{figure}

In this appendix we describe the phase diagram in case $ K<0 $, which is given in~\cref{fig:phase-diagram-K<0}.
For $ J_\perp^2>K^2 $ this system once again only has limit cycles with $ \eta=0 $ and the dynamics of the system is described by~\cref{eq:limit-cycle-mu-sol,eq:precessional-frequency}.
On the other hand, when $ K^2<J_\perp^2 $, two types of limit cycles are stable if $ I_s>I_{s,c} $, as we have seen in~\cref{app:further-limit-cycle,app:stability-eta-neq-0,sec:non-linear-dynamics}.
One of these limit cycles has $ \eta=0 $ and is described by~\cref{eq:limit-cycle-mu-sol,eq:precessional-frequency}.
The stability requirement $ |J_\perp|(1+\mu^2)>K(1-\mu^2) $ for $ K<0 $, in this case leads to $ (B-I_s/2\alpha)^2<(K-J_\perp)^3/(J_\perp+K) $.
The second type of limit cycle has $ \eta\neq0 $ is discussed in~\cref{app:further-limit-cycle} and only exist for current larger then~\cref{eq:critical-current-histeretic-regime}.
We denote the region in~\cref{fig:phase-diagram-K<0} in which both limit cycles exist as the hysteretic regime.

The regions of stable static configurations with $ \eta=1 $ and $ \mu=0 $ are given by~\cref{eq:fixed-point-mu-eta=0,eq:stability-requirement-static-appendix}. Additionally, we find the region of stability with $ \mu=\pm1 $ and $ \eta=0 $ by requiring $ \partial_\mu\dot\mu<0 $ on~\cref{eq:fixed-point-mu-eta=0}.
This results in $ B-I_s/2\alpha<\mp(K-|J_\perp|) $.

\end{document}